\documentstyle[twocolumn,aps,prb,epsf]{revtex}

\begin{document}

\title{Molecular dynamics of folding of secondary structures in Go-type
models of proteins}

\author{Trinh Xuan Hoang and Marek Cieplak}

\address{Institute of Physics, Polish Academy of Sciences, 
Aleja Lotnikow, 02-668 Warsaw, Poland}

\vskip 20pt

%\newpage

\address{
\centering{
\medskip\em
{}~\\
\begin{minipage}{14cm}
We consider six different secondary structures of proteins and construct
two types of Go-type off-lattice models: with the steric constraints and
without.  The basic aminoacid-aminoacid potential is Lennard Jones for
the native contacts and a soft repulsion for the non-native contacts.
The interactions are chosen to make the target secondary structure be
the native state of the system.
We provide a thorough equilibrium and kinetic characterization of the
sequences through the molecular dynamics simulations with the Langevin noise.
Models with the steric constraints are found to be better folders and to
be more stable, especially in the case of the $\beta$-structures.
Phononic spectra for vibrations around the native states
have low frequency gaps that correlate with the thermodynamic stability.
Folding of the secondary structures proceeds through a well defined
sequence of events. For instance, $\alpha$-helices fold from the ends first.
The closer to the native state, the faster establishment of the contacts.
Increasing the system size deteriorates the folding characteristics.
We study the folding times as a function of viscous friction and 
find a regime of moderate friction with the linear dependence.
We also consider folding when one end of a structure is pinned
which imitates instantaneous conditions when a protein is being synthesized.
We find that, under such circumstances, folding of helices is
faster and of the $\beta$-sequences slower.
%{}~\\
%{}~\\
%{\noindent PACS numbers: 87.15.By, 87.10.+e}
\end{minipage}
}}

\maketitle

\section{INTRODUCTION}
Understanding of the statistical mechanics aspects of protein folding
has been recently advanced through studies of coarse grained models in
which aminoacids are represented by single beads. In particular, many
valuable insights have been gained by considering such models on a
lattice (see, e.g.  ref.\cite{Chan,Dill,Cieplak}).  These toy models
have allowed  one to relate the folding process to the sequence
dependent energy landscapes,\cite{Chan1,Wolynes,Sali} to identify the
folding pathways,\cite{Chan2} and to study the issues of
designability.\cite{Tang,Vendrus} They have also been used to
demonstrate existence of tree like kinetic connectivities in the
folding funnel \cite{Garst} that can be represented with the use of
the disconnectivity graphs.\cite{Becker}  

More realistic coarse grained models, however, require an off-lattice
setting.  Recently, there have been a number of off-lattice studies,
which focus on the kinetics of folding,
\cite{Parisi,Irback1,Thirumalai,Klimov1,Stanley} as well as on the
sequence design and determination of the interaction potentials.
\cite{Irback2,Clementi} It has been also suggested that the geometry
of the native structure of proteins itself, without a detailed
information regarding the amino acid sequences, plays a decisive role
in the folding process.\cite{Maritan}  The effective interactions
between the beads in the coarse-grained models are difficult to derive
from microscopic considerations and instead they may be chosen to
reflect statistical properties of protein structures as collected in
the protein data bank. \cite{Miyazawa,Kolinski} 

The advantage of the lattice models is that they allow for an 
enumeration of conformations, at least for short chains, and thus for
identification of the native state and  determination of equilibrium
properties of the system. However, the dynamics of these models
are  not related to any Newton's equations (in the classical
limit) since they have to be defined in terms of the discrete Monte Carlo
steps made within a declared set of allowed moves.

Molecular dynamics (MD) simulations are a natural tool to study models
set in a continuum space independent of whether they are
coarse-grained or fully atomic.  A decade ago, MD simulations of the
microscopic representation of a polypeptide chain could explore time
scales which were around nanoseconds and thus orders of magnitude too
short to study the full duration of a typical folding.\cite{Karplus}
Currently, microsecond time scales have become accessible
\cite{access} (in a special purpose computer) but these feats allow
one to monitor individual trajectories in very restricted regions of
the phase space.  A detailed characterization is still restricted to
up to 10$ns$ long time scales.\cite{Boczko}  Thus such chemically
realistic models cannot yet provide a sufficiently thorough
equilibrium and kinetic characterizations of the system that are
expected when setting up a model to be studied within the framework of
statistical mechanics.

Thus there is a need to study simplified coarse-grained continuum
space models to understand the generic features of folding.  These
models must involve idealized potentials, for instance of the
Lennard-Jones kind.  These interactions may either be constrained to
correspond to a target native state or they are not.\cite{Parisi} The
targeting may be facilitated by augmenting the model by an
introduction of steric constraints,\cite{Thirumalai,Scheraga} which
takes into account the properties of the peptide bonds, but it can
also be accomplished without such constraints.\cite{Li2} The dynamics
of the simplified Lennard-Jones models are usually studied by the
methods of MD. A novel and efficient variant of the MD technique has
been recently proposed by He and Scheraga,\cite{Scheraga} in which one
focuses the evolution on the torsional degrees of freedom.  Some Monte
Carlo studies for those models are also available.
\cite{Parisi,Irback1,Li2} The applicability of the Monte Carlo methods
to dynamical properties (as opposed to equillibrium) remains, however,
untested.

In this paper, we report on molecular dynamics studies of possibly the
simplest models with the native states defined by target conformations --
the off-lattice versions of the Go models.\cite{Go} The interesting
property of the Go model is that it essentially avoids the issue of the
correct specification of the aminoacid-aminoacid interaction and yet it
corresponds to a realistic conformation by defining the couplings in terms
of the target conformation.  Specifically, we consider the Lennard-Jones
interactions between the beads and make them attractive for native contacts
(two non-contiguous beads form a contact if they do not exceed a certain
cutoff distance) and repulsive for non-native contacts. Such models are
protein-like also in the sense that they minimize the structural
frustration.  The object of our studies here is to investigate how viable
are such Lennard-Jones-Go models in the context of the kinetics of protein
folding. Our motivation for performing these studies follows
from the expectation that such models may play a role similar to that of
the Ising spin models in representing properties of the more complicated
real life magnetic systems.

Recently, there have been several related studies of the coarse-grained
off-lattice Go models which, however, asked different questions than in
this paper and were not based on the Lennard-Jones potentials.  The study
by Zhou and Karplus,\cite{Zhou} employed a square well potential (also used
before to analyze homopolymers \cite{Hall}) which leads to a simplified
discrete MD treatment. The authors have studied possible scenarios  (with
or without long-lasting intermediates) of the folding kinetics in a
three-helix-bundle-like protein model as a function of the strength of the
non-native contacts relative to the strength of the native ones.  This kind
of discrete MD have been also used by Dokholyan, Buldyrev, Stanley and
Shakhnovich \cite{Stanley} to identify a folding nucleus in a Go-type
heteropolymer.  (For a related full atom study of this issue for the CI2
protein see ref. \cite{Daggett}).  Another study, by Hardin, Luthey-Schulten,
and Wolynes \cite{Hardin} used a detailed backbone representation   and
implemented an associative memory Hamiltionian in which the contact
potentials are set to be the Gaussian functions. This study was aimed at
understanding the kinetics of the secondary structures formation from the
perspective of the energy landscape picture.  Our partiality to the
Lennard-Jones potentials is of a twofold nature.  First, these potentials
are well established in simulations of liquids and --essentially
--continuum time stable MD codes are available. Second, their overall
distance dependence seems qualitatively correct on a fundamental level.

Real life conformations of proteins in the native state consist of
interconnected secondary structures such as $\alpha$-helices,
$\beta$-hairpins (see, e.g. ref.\cite{Blanco}), and $\beta$-sheets.
Understanding of the kinetics of protein folding should be first
accomplished at the level of the secondary structures.  This is, in fact,
the task of this paper and we narrow it to two classes of the Go models:
with and without the steric constraints and for each of these we consider
six possible secondary structures with different numbers of monomers.  The
kinetics of folding are studied by the standard techniques of MD with a
Langevin noise which controls the temperature of the systems and, at the
same time,  mimics the interactions of the protein fragments with the
molecules of water.

Our method of constructing the off-lattice Go models is outlined in Section
II.  In Section III, we determine the sizes of the native basins of the
models through the shape distortion method \cite{Li} instead of making
assumptions about them, as it is done usually.  In Section IV, we determine
the phononic spectra of the models and discuss their relationship to the
thermodynamic stability of the native structures.  In Section V, we
determine the folding times, the thermodynamic stability and other
characteristic parameters for the models without the steric constraints and
find that the $\beta$-structures have very low thermodynamic stability in
this model. The folding times vary with the native conformation and the
chain length.  In particular, the $\alpha$-helix folds faster than the
$\beta$-hairpin of the same length. The latter also folds much slower than
a similarly sized $\beta$-sheet with three strands.  Experiments on folding
indicate that the folding time for the hairpin is about 30 times slower
that that for the $\alpha$-helix.\cite{Munoz} Our results do not yield the
same rate but, at least, they show that the $\beta$-hairpin is the slower
folder. A similar result has been obtained by Hardin, Luthey-Schulten,
and Wolynes.\cite{Hardin}
Our studies of the three sizes of the helices confirm the general
observation that folding properties deteriorate with the growth of the
system size.\cite{scaling} In Section VI, we repeat the MD studies for the
models with the steric constraints and find that these constraints
substantially improve the thermodynamic stability, especially for the
$\beta$-structures.  At the same time, however, they raise the temperature
of the onset of the glassy effects so the net result is that the
$\alpha$-helices remain good folders and the $\beta$-structures remain bad
folders but  their foldability is improved.  Notice, however, that our
models allow for good or adequate folding, depending on the system, without
any additional dipole-dipole interactions as introduced by He and
Scheraga.\cite{Scheraga}

In Section VII, we focus on the mechanisms of folding and discuss
sequencing of events that takes place in folding of the secondary
structures.  In particular, we investigate characteristic time separations
between the folding steps.  In general, the time separations may depend on
details of a model but the sequencing is expected to be model independent,
i.e. it should proceed as predicted by our Go-type models.  In Section
VIII, we study the dependence of our results on the strength of viscous
damping in the Langevin noise and find that it affects the folding time but
it does not affect any of the other characteristic parameters except when
the friction is set unrealistically low.  Finally, in Section IX, we
consider the processes of folding occurring when one end of the structure
is fixed in space which imitates conditions found when a protein is being
synthesized.\cite{Streyer} We find that fixing one end of a helix
accelerates folding and improves overall folding characteristics.  The
folding is found to originate at the unclamped end.  On the other hand,
folding of the $\beta$ structures  becomes worse when one end is clamped.

This paper sets the stage for studies of full proteins and kinetic interplays
between various secondary structures. Results of such studies for the
Go models of proteins will be presented in a separate paper.

\section{MODELS}

We represent a polypeptide in a simplified manner: by a chain of connected
beads.  The beads' positions correspond to the locations of the C${\alpha}$
atoms.  When the system is in its native state, the beads have no kinetic
energy and the locations of the C${\alpha}$ atoms can be obtained from the
Protein Data Bank. However, the secondary structures that we study are
idealized and are not targeted to a specific real structure. Instead, they
are just meant to be constructed in a way which is very close to typical
$\alpha$-helices and $\beta$-sheets found in the native states of real
proteins.  These secondary structures  are stabilized primarily by the
hydrogen bonds.  The structures that we study are shown in Figure 1.  These
are: three $\alpha$-helices, denoted as H10, H16 and H24, two
$\beta$-hairpins, denoted as B10 and B16, and a $\beta$-sheet, B15. The
numbers in the labels indicate  the numbers of beads in the structures.

All of the bonds between two consecutive beads along the chain of the
helices, i.e. the peptide bonds, are assumed to have the same length of
$d_0=3.8\AA$ which is a typical real life value.  As one proceeds along the
axis of the helix (the $z$-coordinate) from one bead to another, the bead's
azimuthal angle is rotated by $100^o$ and the azimuthal length is displaced
by 1.5$\AA$ which again corresponds to a typical geometry.

Each of the $\beta$-hairpins, B10 and B16, has two anti-parallel strands
which are connected by a turn. In the B15 $\beta$-sheet there are three
strands and two turns.  In the $\beta$-structures, the strands are not
straight lines but have a zigzagged geometry as shown in Figure 1. The bond
lengths between two connected beads are equal to 3.8$\AA$ but the
displacement along the strand direction (the $z$ axis in Figure 1) is
3.5$\AA$.  The distance between two opposite beads in two bonded strands,
is set to 5$\AA$, which is roughly equal to the hydrogen bond's length.
The turn region is constructed so that the bond length and the zigzag
pattern match.

The potentials of interactions between pairs of the beads are constructed
in a way that ensures that the target structure coincides with the ground
state of the system. i.e. with the native state.  We pick the pair
potentials to be of the Lennard-Jones type and select the parameters in a
Go-type fashion\cite{Go} so that significant attraction is associated with
the native contacts and the non-native contacts are purely repulsive.  In
our model, we assume that a native contact is present if the distance
between the two monomers in the designed structure is shorter than
$7.5\AA$. 

\begin{figure}
\epsfxsize=3.2in
\centerline{\epsffile{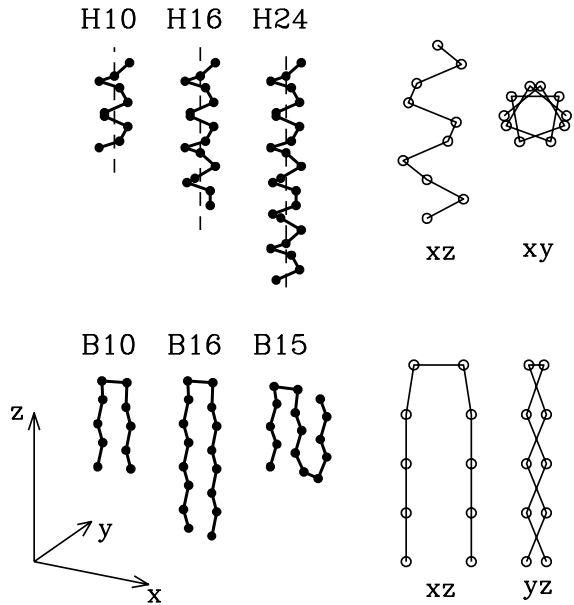}}
\caption{Stereographic projections of the target structures studied in this
paper: three $\alpha$-helices H10, H16, H24, two $\beta$-hairpins B10, B16,
and a $\beta$-sheet B15.  The $xz$ and $yz$ planar projections for H10 and
B10 are shown on the right hand side.  }
\end{figure}

These Go-type couplings already stabilize the structures under studies but
that stabilization will be found here not to be sufficiently adequate,
especially in the case of the hairpins. We thus consider two classes of
models: with and without additional steric constraints.  The steric
constraints add extra stability and they take into account the directional
character of the peptide bonds in a more realistic manner.

\subsection{Go-type model with no steric constraints}

This model is similar in spirit to that introduced by Iori, Marinari, and 
Parisi\cite{Parisi} and to that studied by Li and Cieplak.\cite{Li2}
The conceptual difference between these two papers is that the
former is not constructed in reference to any predetermined target structure.
For a conformation defined by the set of position vectors
$\{ {\bf r}_i \}$, $i=1,2 \ldots N$, the potential energy
is assumed to take the following form
\begin{equation}
E_p(\{{\bf r}_i\}) =  V^{BB} + V^{NAT} + V^{NON} \;\;.
\label{eqep}
\end{equation}
The first term represents rigidity of the backbone potential, the second
term corresponds to interactions in the native contacts and the third term
to those in the non-native contacts. $N$ denotes the number of residues.

The backbone potential takes the form of the sum over harmonic\cite{Parisi}
and anharmonic\cite{Clementi} interactions
\begin{equation}
V^{BB} = \sum_{i=1}^{N-1} [ k_1 (r_{i,i+1} - d_0)^2 +
k_2 (r_{i,i+1}-d_0)^4],
\end{equation}
where $r_{i,i+1}=|{\bf r}_i - {\bf r}_{i+1}|$ is the distance between
two consecutive beads; $d_0 = 3.8 \AA$, $k_1=\epsilon$ and
$k_2=100\epsilon$, where $\epsilon$ is the  Lennard-Jones
energy parameter corresponding to a native contact.

The interaction between residues that form a native contact in
the target structure is taken to be of the Lennard-Jones form:
\begin{equation}
V^{NAT} = \sum_{i<j}^{NAT}4\epsilon \left[ \left( \frac{\sigma_{ij}}{r_{ij}}
\right)^{12}-\left(\frac{\sigma_{ij}}{r_{ij}}\right)^6\right],
\end{equation}
where the sum is over all pairs of residues $i$ and $j$ 
(but those which are immediate neighbors along the chain) which
form the native contacts in the given target structure.
$r_{ij}=|{\bf r}_i-{\bf r}_{j}|$ is the monomer to monomer
distance.  The parameters $\sigma_{ij}$ are chosen in a way that
each contact in the native structure is stabilized 
at the minimum of the potential.
Essentially, $\sigma_{ij}=2^{-1/6}\cdot d_{ij}$,
where $d_{ij}$ is the corresponding native contact's length.

Residues not forming the native contacts interact via a repulsive soft core
potential. Our potential for non-native contacts, given below, differs from
the model of Iori {\it et al.} in that it falls to $0$ after some cut-off
distance, $d_{cut}$.  The purpose of the cutoff is to minimize structural
frustration and thus to improve foldability.

\begin{equation}
V^{NON} = \sum_{i<j}^{NON} V_{ij}^{NON},
\end{equation}
\begin{equation}
V_{ij}^{NON} =
\left\{
\begin{array}{ll}
4\epsilon \left[ \left( \frac{\sigma_0}{r_{ij}}\right)^{12}
- \left( \frac{\sigma_0}{r_{ij}}\right)^{6} \right]+ \epsilon &, r_{ij} <
  d_{cut}\\
0 &, r_{ij} \geq d_{cut}.
 \end{array}
\right.
\end{equation}
Here, $\sigma_0=2^{-1/6} \cdot d_{cut}$.
For distances shorter than $d_{cut}$ the potential is
purely repulsive. For the $\alpha$-helices we chose 
$d_{cut}=\left<d_{ij}\right> \approx 5.5\AA$
which seems to be a natural choice. For the $\beta$-structures, however,
such a choice of $d_{cut}$ leads to an instability of the native
conformation due to a substantial energy degeneracy of the nearby
conformations. A sufficient extending of the cutoff  removes the degeneracy
and stabilizes the native state. We have found that an adequate choice for
the $\beta$-structures is to take $d_{cut}=7.5\AA$ -- the distance that
defines what is a contact.

Notice that in the target structures, all native contacts are optimized
while the non-native contact potentials give no contribution to the energy.
Thus the target structure is indeed the native state of the system. The
non-native interactions contribute only when moving away from the target.

\subsection{Go-type model with the steric constraints}

In this case, the potential energy of
the system is given by
\begin{equation}
$\~E$_p(\{{\bf r}_i\}) =  V^{BB} + V^{NAT} + V^{NON} + V^{BA} + V^{DA},
\end{equation}
where the first three terms are identical to those in equation (\ref{eqep}), 
whereas the last two
terms correspond to the bond angle and dihedral angle potentials
respectively. The bond angle, $\theta_i$, is defined as the angle between two
successive vectors ${\bf v}_{i}$ and  ${\bf v}_{i+1}$, where
${\bf v}_{i} = {\bf r}_{i+1} - {\bf r}_i$. The dihedral angle, $\phi_i$,
is the angle between 
% two planes defined by the 
two vector products 
${\bf v}_{i-1} \times {\bf v}_{i}$ and ${\bf v}_{i} \times {\bf v}_{i+1}$.
Following ref. \cite{Thirumalai}, we use the following potentials for the bond
and the dihedral angles
\begin{equation}
V^{BA} = \sum_{i=1}^{N-2} \frac{k_\theta}{2} (\theta_i - \theta_{0i})^2
\end{equation}
\begin{equation}
V^{DA} = \sum_{i=1}^{N-3} [ A(1+cos\phi_i)+B(1+cos3\phi_i)]
\end{equation}
where $k_\theta=20\epsilon/(rad)^2$, $A=0\epsilon$ and $B=0.2\epsilon$.
Our angle dependent potentials differ from those in ref. \cite{Thirumalai}, since in our case
we take
$\theta_{0i}$ to be, in general, site-dependent and equal to the bond angles
in the native targets. For the helices, however, $\theta_{0i}$ are
uniform and equal to 1.56157 $rad$ (89.4714 $deg$).  For the $\beta$-structures,
$\theta_i$'s in the turn region are different from those in the strands.

Introduction of the steric constraints to the model described by eq. (1)
shifts the native state away from the target conformation because the
target need not correspond to a minimum of the dihedral potentials.
However, we have found that for our choice of the parameters $A$ and $B$
the true native states differ only little from the targets.  The true
native states are found by a multiple zero temperature quench procedure
from low energy conformations generated by MD trajectories that start in
the target conformation.  The conformation distances from the native states
to the targets (to be defined in Sec. III) never exceeded
$0.05\AA$.

It should be pointed out that none of the models studied here
can distinguish between the right- and left-handed helices.
The $\alpha$-helices found in nature, however, are only right-handed.
The model in which only right-handed helices are favored should include
terms related to the chirality of the chain conformation. The chirality
can be given by 
$sign[({\bf v}_{i-1}\times {\bf v}_i)\cdot {\bf v}_{i+1}]$, where 
a negative sign means a left-handed conformation and a positive sign
means a right-handed conformation.\cite{Kolinski}

In the following Sections, the Go-type sequences without the steric
constraints will be denoted by the symbols associated with the targets:
H10, H16, H24, B10, B16, B15.  On the other hand, the sequences constructed
with the additional steric constraints will be labeled as \~H10, \~H16,
\~H24, \~B10, \~B16, \~B15.

\subsection{Dynamical equations and the thermostat}

The motion of the model secondary structures of proteins can be described 
by the Langevin equation
\begin{equation}
m\ddot{{\bf r}} = -\gamma \dot{{\bf r}} + F_c + \Gamma
\label{eqlang}
\end{equation}
where ${\bf r}$ is a generalized coordinate of a bead, $m$ is the monomer's
mass, $F_c=-\nabla_r E_p$ is the conformation force, $\gamma$ is a friction
coefficient and $\Gamma$ is the random force which is introduced to balance
the energy dissipation caused by friction.  Both the friction and the
random force represent the effects of the solvent and they control the
temperature.\cite{Grest} $\Gamma$ is assumed to be drawn from the Gaussian
distribution with the standard variance related to temperature by
\begin{equation}
\left<\Gamma(0)\Gamma(t)\right> = 2\gamma k_B T \delta(t),
\label{eqgam}
\end{equation}
where $k_B$ is the Boltzmann constant, $T$ is temperature, $t$ is time
and $\delta(t)$ is the Dirac delta function.

The Langevin equations are integrated using the fifth order
predictor-corrector scheme.\cite{Allen}  The friction and random force
terms are included in the form of a noise perturbing the Newtonian motion
at each integration step.  In the case of the model with the steric
constraints, the forces associated with the angle-dependent potentials are
calculated through a numerical determination of the derivatives of the
potential.

In the following, the temperature will be measured in the reduced units of
$\epsilon/k_B$.  The integration time step is taken to be $\Delta t = 0.005
\tau$, where $\tau$ is a characteristic time unit.  At low values of
friction, $\tau$ coincides with the period of oscillations, $\tau _m$ near
the Lennard-Jones minimum and is equal to $\sqrt{m a^2/\epsilon}$, where
$a$ is a Van der Waals radius of the amino acid residues. The value of $a$
is chosen to be equal to $5\AA$, and this value is roughly equal to
$\left<\sigma_{ij}\right>$.
As estimated in ref. \cite{Thirumalai}, the typical values of $m$ is
$3\times 10^{-22}g$ and $ \epsilon$ is of the order of $1kcal/mol$, hence
$\tau$ is roughly equal to $3ps$. The period $\tau _m$, however, depends on
the friction coefficient, $\gamma$, as discussed in ref.
\cite{Thirumalai,Klimov1}.  Most of our simulations are performed with
$\gamma=10 m\tau^{-1}$ which is 5 times higher than a standard choice in
molecular dynamics studies of liquids.  Higher values of $\gamma$ may be
actually more realistic.  For a single aminoacid in water, an effective
$\gamma$ has been argued to be even 
of order $50 m /\tau$.\cite{Thirumalai,Klimov1} However, the MD code
becomes unstable for too high $\gamma$'s.  The dependence of the folding
characteristics on $\gamma$ will be discussed later in this study.

\section{DETERMINATION OF THE NATIVE BASIN}

One problem that is encountered when studying off-lattice models is
providing a definition of what small distortion away from the native state
can still be considered as belonging to the native state.  In other words -
what is the definition of the native basin (which should not to be confused
with the folding funnel)? Answering this question is important when
calculating the folding time (is the system already in the native basin?)
and when calculating the equilibrium stability of the basin.  A need for
delineation of the native basin does not arise in lattice models due to the
discretization of the possible shapes -- there, it is just one
conformation.

The delineation of the native basin can be accomplished by considering the
conformational distance, $\delta$, away from the native state. $\delta$ is
given by 
\begin{equation}
\delta^2=\frac{2}{N^2-3N+2}\sum_{i=1}^{N-2}\sum_{j=i+2}^N
(r_{ij}-r_{ij}^{NAT})^2,
\end{equation}
where $r_{ij}$ and $r_{ij}^{NAT}$ are the
monomer to monomer distances in the given structure and in the native state
respectively. The distance $\delta$ is measured in $\AA$.

\begin{figure}
\epsfxsize=3.2in
\centerline{\epsffile{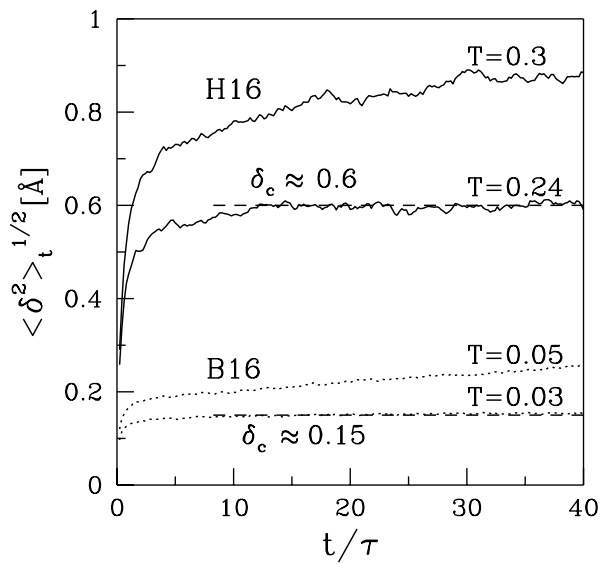}}
\caption{The average root mean square
distance to the native state as a function of time when the sequence is
placed in its native conformation. The figure is for sequences H16 and B16
and for the reduced temperatures as indicated.
The results are averaged over 500 trajectories which differ in the
strings of the random noise.}
\end{figure}

We estimate the characteristic size of the native basin by using the
shape distortion method proposed in ref. \cite{Li}. The method is based on
monitoring the conformational distance to the native state as a
function of time and it considers trajectories that originate from the
native state. At a sufficiently large time scale the conformational
distance saturates below some critical temperature, $T_c$. The
saturation value of the distance at this temperature, $\delta _c$, is
used  for the estimated native basin's size and the results are found
to be consistent with the quench-obtained sampling of the local energy
minima of the system. At the same time, $T_c$ has been found\cite{Li}
to be a measure of the folding temperature $T_f$. $T_f$ itself is
defined as a temperature at which the equilibrium probability of
finding the system in its native basin crosses 1/2.

An illustration of the shape distortion method for sequences H16 and
B16 is shown in Figure 2.  For each of these sequences the dependence
of $\left<\delta^2\right>_t^{1/2}$ on time is shown for two
temperatures: a critical $T_c$, when the saturation is still observed,
and a somewhat higher $T$, when there is no saturation which is
interpreted as exiting the ``trap'' provided by the native basin.  The
estimated basin size for sequences H16 and B16 are $0.6\AA$ and
$0.15\AA$ respectively. Notice that sequence H16 has much larger basin
than sequence B16 and the corresponding $T_c$ is also much higher for
H16. 

Table I summarizes the values of $\delta_c$ and $T_c$ for all of the
sequences studied. It also shows the number of native contacts in the
native state, $N_c$, the conformational distance from the native state
to the closest local minimum, $\delta_1$, and the values of $T_f$
obtained in the next Sections.  The closest local minimum to the
native state is found by a multiple quenching from the random
conformations in a low $T$ MD trajectory.  Notice that for the
$\beta$-structures, the models with the steric constraints yield a
larger $\delta _c$ compared to the models without the constraints.
The basin sizes for the $\alpha$-helices, on the other hand, are
comparable.  In most cases, we have found that $\delta _c$ is somewhat
smaller than $\delta _1$. This is not so, however, in the case of
\~H24 and \~B15. We interpret this as a necessity to delineate the
native basin by more than one isotropic parameter $\delta_c$ and yet
we shall use it here as a simple way to provide estimates.

\vbox{
\begin{table}[t]
\begin{center}
\begin{tabular}{ c c c c c c c}
SEQ & $N_c$ & $\delta_1 [\AA]$ & $\delta_c [\AA]$ & $T_c$ & $T_f$&$T_{min}$  \\
\hline
H10   & 21 & 0.8248 & 0.50(3) & 0.23(1)  & 0.24(1)  & 0.25(3) \\
H16   & 39 & 0.6594 & 0.60(3) & 0.24(1)  & 0.24(1)  & 0.30(3) \\
H24   & 63 & 0.5430 & 0.53(3) & 0.19(1)  & 0.20(1)  & 0.30(3)  \\
B10   & 18 & 0.2588 & 0.15(2) & 0.035(2) & 0.033(2) & 0.08(1) \\
B16   & 33 & 0.2288 & 0.15(2) & 0.030(2) & 0.027(2) & 0.07(1) \\
B15   & 33 & 0.1382 & 0.13(2) & 0.030(2) & 0.029(2) & 0.05(1) \\
\hline
\~H10 & 21 & 0.8133 & 0.46(3) & 0.30(1)  & 0.31(1)  & 0.50(5) \\
\~H16 & 39 & 0.6356 & 0.55(3) & 0.30(1)  & 0.29(1)  & 0.45(3) \\
\~H24 & 63 & 0.3896 & 0.55(5) & 0.24(2)  & 0.24(1)  & 0.36(3) \\
\~B10 & 18 & 0.3178 & 0.25(2) & 0.20(1)  & 0.20(1)  & 0.35(5) \\
\~B16 & 33 & 0.2801 & 0.25(2) & 0.20(1)  & 0.21(1)  & 0.35(2) \\
\~B15 & 33 & 0.2045 & 0.30(2) & 0.22(1)  & 0.20(1)  & 0.35(2) \\
\end{tabular}
\end{center}
\caption{The number of native contacts in the native conformations,
$N_c$, the conformational distance from the native state to the
closest local minimum, $\delta_1$, the native basin size, $\delta_c$,
the critical temperature of the shape distortion, $T_c$, the folding
temperature, $T_f$, and the temperature of the fastest folding,
$T_{min}$, for the sequences studied.
The numbers in parenthesis indicate the error bars.}
\end{table}
}

\section{PHONONIC SPECTRA AND STABILITY}

We now consider elastic vibrations of the systems around their
native states and ask if the phononic spectra relate to the folding
temperature.

Let $u_{i\alpha}$ denote a small displacement of bead $i$
in the direction $\alpha$ ($\alpha=x,y,z$) from its native position.
In the harmonic approximation,
the motion of the system is governed by a set of $3N$ coupled 
equations\cite{Callaway}
\begin{eqnarray}
- m \omega^2 u_{i\alpha} = \sum_{j=1}^N \sum_{\beta} 
k^{\alpha\beta}_{ij} \; u_{j\beta} \;\; ,  
\end{eqnarray}
where $\omega$ is an angular frequency, and $k^{\alpha\beta}_{ij}$
are the second derivatives of the potential energy taken at the native
conformation:
\begin{equation}
k^{\alpha\beta}_{ij} = \left. \frac{\partial^2 E_p}{\partial u_{i\alpha}
\partial u_{j\beta}} \right|_{{\bf u}=0} \; .
\end{equation}
We diagonalize these equations with the use of the Jacobi transformation
method\cite{NR} and determine the phononic spectrum from the eigenvalues.
The elastic constant matrix $\{k^{\alpha\beta}_{ij}\}$ is real and
symmetric.  For the models without the steric constraints, when only the
pair-wise interactions are present, the diagonal elements are equal to the
negative of the sum over all of the off-diagonal elements in the same row
(or column).  In the case of the models with the steric constraints,
however, the potentials acquire many-body components.  The elastic
constants $k^{\alpha\beta}_{ij}$ are calculated by freezing all beads but
one, at a time, in their native positions and by  measuring the resulting
increase in the force when the unfrozen bead is displaced in a given
direction. This is done numerically and then a displacement of $10^{-7}\AA$
has been found to yield a sufficient accuracy. The elastic constants, when
measured along a line connecting the two beads, are of order of
$2\epsilon/\AA^2$ for both the contact and peptide bond interactions.  The
steric constraints enhance the elastic constants by up to a factor of 2.
Naturally, this will increase the stability.

The phononic spectra for the models without and with the steric constraints
are shown in Figures 3 and 4 respectively.  Each spectrum contains 6 zero
frequency modes which correspond to the translational and rotational
degrees of freedom of the system as a whole.  The first excited mode, of
frequency $\omega _1$, defines the frequency gap in the spectrum.  We
observe that the gap correlates with the thermodynamic stability:
generally, the bigger the $T_f$, the bigger the gap.  Specifically, we
notice that, for a given kind of structure, $\omega _1$ decreases as the
system size increases.  Furthermore, the gaps for the $\alpha$-helices are
larger than those for the $\beta$-structures for a given chain length.
Finally, adding the steric constraints shifts the whole spectra towards
higher frequencies. 

An energy associated with a mode depends on its amplitude. If an amplitude
of vibrations of a single bead is of order $u_0$, then the energy of the
first excited mode, $E_1$, is of order $N m \omega_1^2 u_0^2$. As an
estimate of $u_0$, we may take $\approx 0.1a$ which is an analog of the
Lindemann criterion for melting of the solids.  Note that $\omega_1$ is
expressed in units of $1 / \tau$, where $\tau=\sqrt{m a^2/\epsilon}$ and
$a=5\AA$. Thus our estimates for $E_1/\epsilon$ are 1.32, 0.77 and 0.28 for
H10, H16 and H24 respectively.  This indicates a decrease of stability with
$N$ but the correlation with the actual value of $T_f$ is weak.

\begin{figure}
\epsfxsize=3.2in
\centerline{\epsffile{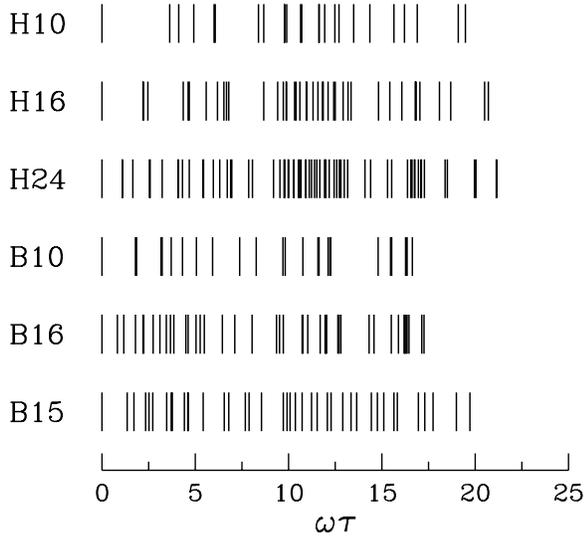}}
\caption{The phononic spectra for the sequences modeled without
the steric constraints. The values of $\omega _1 \tau$ given in $rad$
 are, top to bottom: 3.635, 2.195, 1.080, 1.789, 0.815, 1.339.}
\end{figure}

\begin{figure}
\epsfxsize=3.2in
\centerline{\epsffile{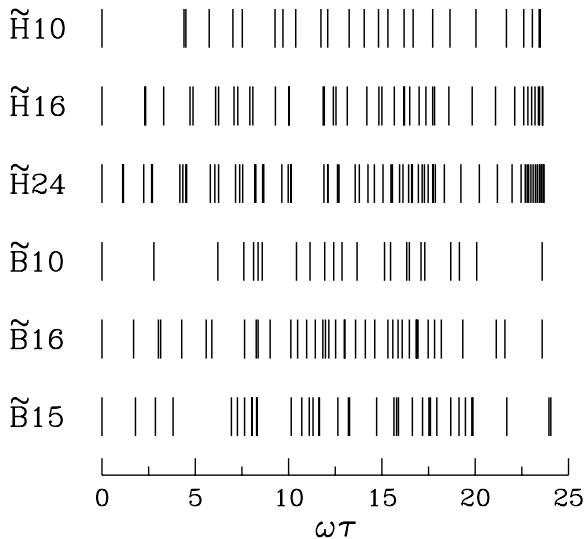}}
\caption{Same as Fig. 3 but for the sequences modeled with the
steric constraints. The values of $\omega _1 \tau$ are, top to bottom:
4.385, 2.290, 1.105, 2.775, 1.695, 1.795.}
\end{figure}

\section{FOLDING PROPERTIES IN MODELS WITHOUT THE STERIC CONSTRAINTS}

\subsection{Equilibrium properties}

Before we discuss the folding process, we first provide a thermodynamic
characterization of the sequences studied.  We focus on three parameters:
${\mathcal P}_0$, $C$, and $\chi$.  These denote the probability of being
in the native basin, the specific heat per bead, and the structural
susceptibility respectively.  The thermodynamic stability may be
characterized by $T_f$ at which ${\mathcal P}_0$ crosses $\frac{1}{2}$.
The specific heat is defined by the energy fluctuations: 
\begin{equation}
C = \frac{1}{N}\frac{\left<E^2\right>-\left<E\right>^2}{T^2} \;\;,
\end{equation}
where $E$ is the total energy (kinetic and potential) of the system.  The
brackets denote the thermodynamic average.  The structural susceptibility
is defined in terms of the structural overlap function\cite{Thirumalai}
which is given by
\begin{equation}
\chi_s=1-\frac{2}{N^2-3N+2}\sum_{i=1}^{N-2}\sum_{j=i+2}^N
\Theta (\delta_c - |r_{ij} - r_{ij}^{NAT}|),
\end{equation}
where $\Theta(x)$ is the Heavyside function. $\chi$ is then a measure of
fluctuations in $\chi _s$: 
\begin{equation}
\chi = \left<\chi_s^2(T)\right> - \left<\chi_s(T)\right>^2
\end{equation}

The maximum in $C$,  when plotted against $T$, corresponds to the collapse
transition temperature $T_\theta$ at which there is a transition from
random coil to compact conformation. The maximum in $\chi$ may be
interpreted as a signature of the folding temperature
$T_f$.\cite{Thirumalai1,Camacho} It has been suggested in ref.
\cite{Klimov2} that a small difference between these two temperatures is
indicative of good folding properties.  As a practical criterion for good
foldability one may take the parameter $\sigma _T=(T_\theta-T_f)/T_\theta$
to be less than 0.4.\cite{Thirumalai}

We calculate the thermodynamic parameters by averaging over many long MD
trajectories using the native state as the starting configuration to maker
sure that the evolution take space in the right parts of the phase space.
For each temperature, the times used for averaging in each trajectory are
between $500$ and $2000\tau$ depending on the chain length.  The first
$1000\tau$ are not taken into account when averaging.  We used as many
trajectories as needed to provide stable characterization.  In practice,
considering of order  50 trajectories have been found to be sufficient.

The values of $T_f$  are listed in Table I. They almost coincide with the
values $T_c$ obtained by the shape distortion method.  One can clearly see
that the stability of the helices is substantially larger than that of the
$\beta$ structures.  The poor stability of the $\beta$-systems is a result
of their flat non-compact native conformation.  There are many competing
local energy minima with very similar structures in the immediate
neighborhood of the native state.  We have checked that an increase in the
off-plane zigzag like departures can bring about a significant boost to the
stability. However, one of the goals of this paper is to deal with
realistic  geometries.  We shall see, in Sec. VI, that the extra
stability that is needed will be provided by the additional steric
constraints. This is also illustrated in Figure 5 for three selected
sequences.

\begin{figure}
\epsfxsize=3.2in
\centerline{\epsffile{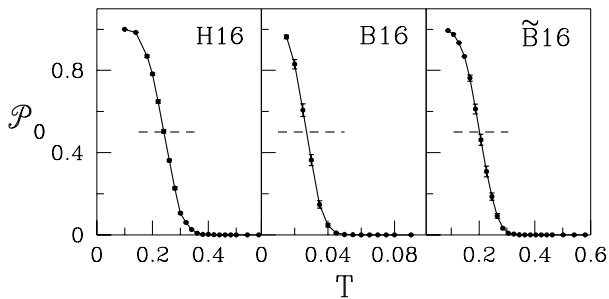}}
\caption{The probability of being in the native basin as function
of temperature for the sequences with $N$=16. H16 and B16 are without
the steric constraints. Adding the steric constraints improves
the stability as seen in the case of \~B16 on the right hand side.}
\end{figure}

\begin{figure}
\epsfxsize=3.2in
\centerline{\epsffile{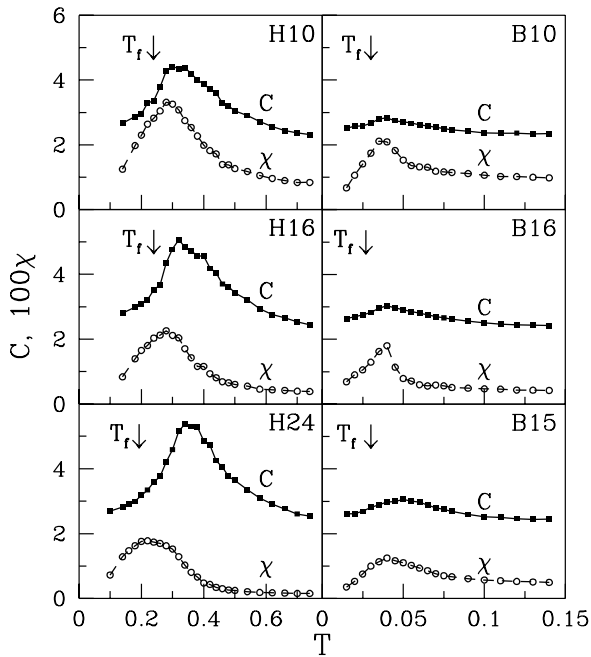}}
\caption{The specific heat, $C$, and the structural susceptibility,
$\chi$, as functions of the reduced temperature for the sequences indicated.
The arrows show the values of the folding temperature as
obtained through monitoring of the equilibrium probability
to reside in the native basin.}
\end{figure}

Figure 6 summarizes the thermodynamic properties of the sequences modeled
without the steric constraints. The peaks of $C$ and $\chi$ are observed
for all of the helices and $\beta$-systems.  Notice, however, that in the
case of the $\beta$-structures the peaks are much less pronounced.  Notice
also that the values of $T_f$, as indicated by the arrows and obtained
based on monitoring ${\mathcal P}_0(T)$, generally agree quite closely with
the positions of the maxima in $\chi$. This consistency testifies about a
reasonable and consistent determination of the size of the native basin.
For all of the sequences studied here the parameters $\sigma _T$ are
smaller than 0.4 and thus are expected to be good folders whereas H24 shows
the borderline behavior.

\subsection{Kinetic properties}

We now consider the kinetic properties of the sequences without the steric
constrains. From a kinetic point of view, a sequence is considered to be a
good folder if its folding temperature $T_f$ is comparable or preferablely
larger than $T_{min}$  -- the temperature of the fastest folding. If this
happens then there must exist a temperature range below $T_f$ at which the
native state is still easily accessible kinetically.  Thus the first task
in this context is to determine $T_{min}$.

We have studied the dynamics of the sequences by extensive MD simulations.
Figure 7 shows the median folding time as a function of temperature.  At
each temperature the median folding time is computed   based on either 100
or 200 trajectories starting from random initial conformations. The initial
conformations are generated from random sets of the bond angles and
dihedral angles, where the bond angles are additionally restricted to have
random value between $0^o$ and $90^o$ so that these conformations are not
compact.  On the other hand, the cases with the distance between two
non-bonded beads of less than $0.8 d_0$ have been also excluded in order
not to violate the self-avoidance condition. Conformations generated by
this method typically look like extended random coils.  Figure 7 shows the
expected U-shape dependence\cite{Socci} of the folding time on temperature.
Notice that the $\beta$-sequences fold the best at a much lower range of
temperatures than what is observed for the $\alpha$-helices. The
temperature of the fastest folding $T_{min}$ for the $\beta$-sequences is
between 0.05 and 0.08 while for the $\alpha$-helices it is about 0.25 or
0.3.

Since for all the sequences $T_f$ is less than $T_{min}$, the sequences can
be considered to be either good or bad folders  depending on how
significantly do the folding conditions deteriorate on moving away from
$T_{min}$ to $T_f$. Thus it is of interest to compare the folding time at
$T_{min}$, $t_1$, to the folding time at $T_f$, $t_2$.  The
$\alpha$-helices are good folders because for them $t_2/t_1 < 2$. i.e., the
two times are almost the same. For the the $\beta$-sequences, on the other
hand, these ratios are higher reflecting a much narrower temperature range
in which the good folding conditions are available. For B10 and B15, for
instance, the ratios $t_2/t_1$ are about 7 and 5 respectively.  Thus the
$\beta$-sequences without the steric constraints are bad folders.
Note that our kinetically derived results on the quality of folding do not
fully agree with those obtained by studying the thermodynamically derived
parameter $\sigma_T$.  The reason why the thermodynamic criterion appears
to be faulty may have to do with the fact that the $\beta$-structures
studied here have a low degree of compactness. Note also the poor
pronunciation of the peaks in the specific heat for the $\beta$-structures.

\begin{figure}
\epsfxsize=3.2in
\centerline{\epsffile{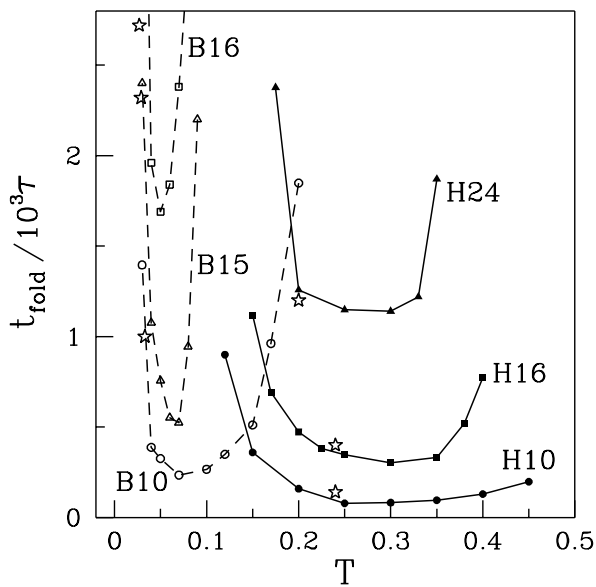}}
\caption{The median folding time versus temperature for the sequences
modeled without the steric constraints. The asterisks indicate
the corresponding values of $T_f$.}
\end{figure}

Figure 7 shows that the $\beta$-sequences fold more slowly than the
$\alpha$-helices of the same chain length. For instance, the minimum
folding time for $\beta$-hairpin B16 is about 5 times larger than for the
$\alpha$-helix H16.  Notice also that the $\beta$-sheet B15 folds
considerablely faster than the hairpin B16, though their lengths are
comparable.  All of these observations are consistent with the large role
of the local contacts (as measured along the chain) in establishing good
foldability.\cite{Unger} 

Figure 7 indicates that the folding time strongly depends on the chain
length. Studies of lattice models, first by Gutin {\it et al.}\cite{Gutin}
and recently by us,\cite{scaling} have pointed out that the folding time
grows as a power law with the system size $t \sim N^{\lambda}$. Our
calculations\cite{scaling} have shown that for the Go sequences $\lambda$
is about 6 for two dimensional lattice sequences and about 3 for three
dimensional sequences. Here, we have checked that the folding times at
$T_{min}$ of the three sequences H10, H16, H24 are consistent by the power
law with $\lambda = 3.0 \pm 0.1$.  The agreement with the exponent obtained
for the maximally compact Go lattice models is probably coincidental.  One
of the important conclusions suggested in ref. \cite{scaling} is that there
may exist a limit to functionality of proteins that can be demonstrated
through studies of scaling of $T_f$ and $T_{min}$. In the Go lattice
models, as $N$ grows, $T_{min}$ grows indefinitely whereas $T_f$ saturates
with $N$. This, together with the increase in the folding times, indicates
a deterioration of the folding properties with $N$.  The data obtained for
the few sizes of helices studied here (see Table I) are only partially
consistent with this scenario for the behaviors of the characteristic
temperatures. Note, however, that the longer a helix, or a
$\beta$-structure, is the less compact is its native conformation and this
in itself may yield a different behavior of $T_f$ and $T_{min}$.

\section{FOLDING PROPERTIES IN MODELS WITH THE STERIC CONSTRAINTS}

\subsection{Equilibrium properties}

We now repeat our methodology in reference to the sequences which are
modeled with the implementation of the steric constraints.  The
thermodynamic properties for the systems of this type are summarized in
Figure 8. The values of the folding temperatures $T_f$ are also listed in
Table I. Notice that the models with the steric constraints are generally
characterized by larger values of $T_f$ than the models without such
constraints. The changes are especially substantial for the
$\beta$-structures since the resulting values of $T_f$ become now
comparable to those of the helices.

\begin{figure}
\epsfxsize=3.2in
\centerline{\epsffile{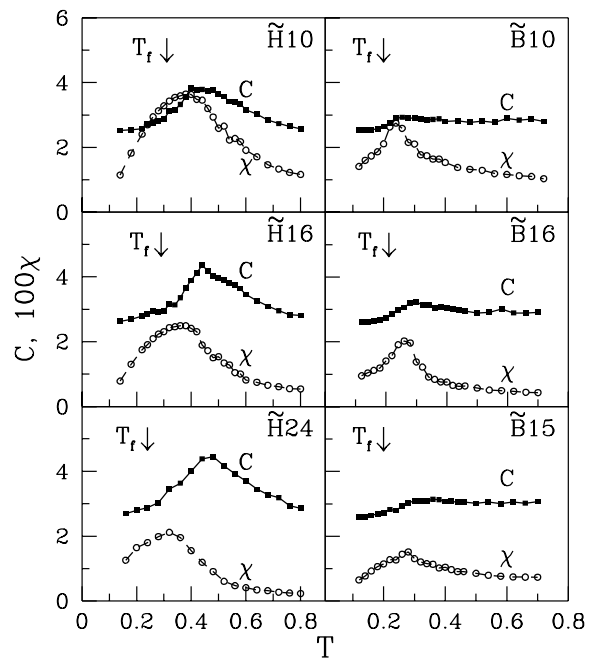}}
\caption{Same as Figure 6 but for the sequences modeled with the
steric constraints. Notice the shift in the temperature scale.}
\end{figure}

\subsection{Kinetic properties}

Figure 9 shows the dependence of the folding time on temperature.  The
points shown are based on either 100 or 200 trajectories starting from
random initial conformations.  For all of the structures studied here,
$T_{min}$  also becomes larger than that in the models without the steric
constraints. The time scales of folding, however, are affected to a
relatively much smaller extent.  For the $\beta$-sequences, the folding
time at $T_{min}$ practically does not change.  For the $\alpha$-helices,
on the other hand, some slowing down is observed.  When comparing the
$\alpha$-helices to the $\beta$-structures one concludes that the helices
continue to be faster folders but, for instance, \~H16 now folds only about
two times faster than \~B16 at $T_{min}$, as opposed to the factor of five
observed without the constraints.

Like in the model without steric constraints, for all the cases we still
have $T_f < T_{min}$.  Comparing the folding times at $T_f$ and at
$T_{min}$ yields that the helices continue to be good folders 
($t_2/t_1 < 2$).  The $\beta$-sequences are still bad folders but now the
ratios $t_2/t_1$ are somewhat smaller than that in the model without steric
constraints.  For instance, for \~B10, $t_2/t_1$ is about 3.   

We conclude that adding the steric constraints to the Go-type interactions
between the aminoacids improves the stability of the sequences, while it
does not change much in the folding characteristics. The folding
characteristics seem to be embedded in the very geometry of the native
structure.  

\begin{figure}
\epsfxsize=3.2in
\centerline{\epsffile{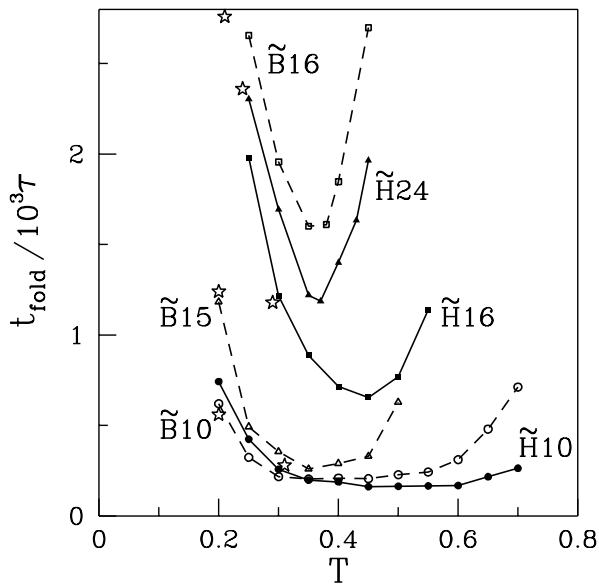}}
\caption{Same as Figure 7 but for the sequences modeled with
the steric constraints.}
\end{figure}

\section{THE MECHANISM OF FOLDING OF SECONDARY STRUCTURES}

\subsection{Sequencing of events}

In this Section, we focus on identifying stages in the folding
process of the model secondary structures. Specifically, we discuss
orders and time scales in which specific contacts are established.
We narrow our discussion now to the three following structures:
the $\alpha$-helix H16, the $\beta$-hairpin B16,
and the $\beta$-sheet B15.

By starting from random unfolded conformations we compute an average time,
at which a given contact is established for the first time.  Two monomers,
$i$-th and $j$-th, are assumed to be in a contact if their distance is
smaller than $1.5\sigma_{ij}$. The contact range definition of $7.5\AA$,
used in the potential design procedure, is not appropriate to be used here,
since each contact has a potential with its own variable $\sigma_{ij}$.
Since the number of all native contacts is quite large we select several
contacts, as indicated in Figure 10, for monitoring.  The choice of the
contacts is motivated by their dominant role in stabilization provided by
the hydrogen bonds.  The labels indicate order in which these selected
contacts first appear on average.  For the $\alpha$-helix, the folding is
seen to start, statistically, at the ends (we have symmetrized the data
with respect to the two ends).  On the other hand both $\beta$-structures
first fold near the turns, i.e.  where the local contacts are present.
Note that the contacts indicated in the helix all have the same locality
index and yet the end contacts are found to be privileged.  Adding the
steric constraints does not change the average sequencing of the kinetic
events.

Figures 10 and 11 show the average times to establish the contacts at
$T_{min}$ for the unconstrained and constrained Go-type models
respectively.  When comparing the times obtained for the $\alpha$-helix to
the $\beta$-structures, note the change in the time scale. The full folding
takes place in about 10 to 20 times longer than the time needed
to establish all contacts.  The general pattern observed here is
that of steps which are initially more or less evenly spaced followed by a
rapid acceleration towards the end.

\begin{figure}
\epsfxsize=3.4in
\centerline{\epsffile{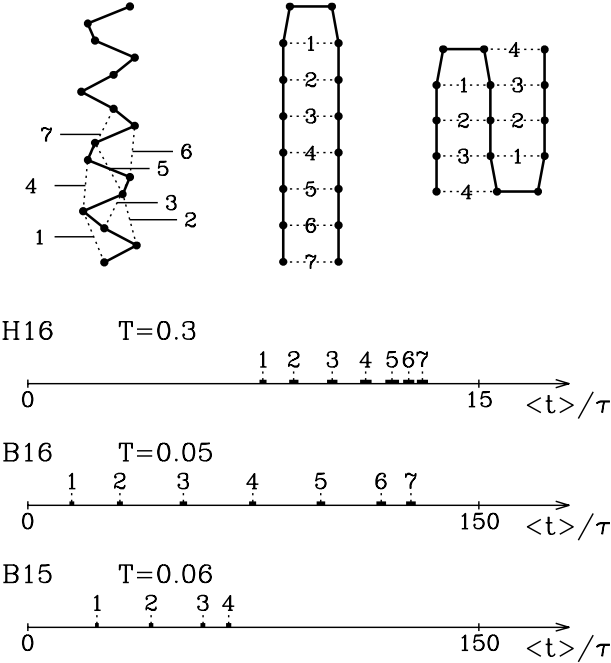}}
\caption{Top: the contacts considered in the sequencing studies together
with the number which indicates the order of the appearance.  The
structures are, from the left to the right, H16, B16 and H15.  Bottom: the
corresponding average time for a contact to appear for the first time
during folding.  The folding takes place at $T_{min}$.  The results for H16
are averaged over 1000 trajectories that start from random conformations.
For B16 and B15, 500 trajectories are considered.  The error bars are
indicated by the thicker horizontal marks.}
\end{figure}

\begin{figure}
\epsfxsize=3.4in
\centerline{\epsffile{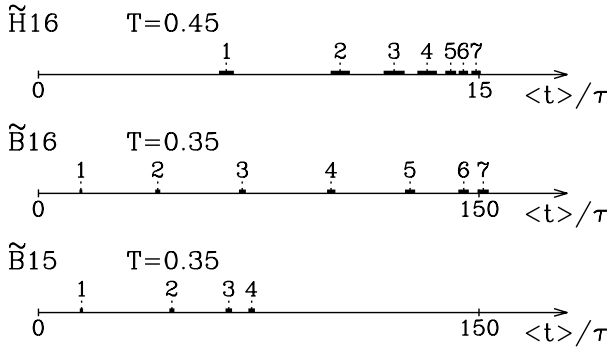}}
\caption{Same as the bottom of Figure 10 but for the sterically constrained
sequences.}
\end{figure}

Figure 12 shows the average times to establish the contacts at temperatures
which are away from $T_{min}$ -- above or below. We focus on \~H16 and
\~B16.  The results show that for sequence \~B16 the sequencing of contacts
does not change with temperature. For \~H16, on the other hand, the
sequencing starts the same as at $T_{min}$, i.e. at the end points, but the
canonical order becomes disrupted at the later stages.  For instance,
establishing contact 5 before 7 is as likely as establishing 7 before 5.
Notice that, above $T_{min}$, establishing the basic contacts in the two
structures takes about as long as at $T_{min}$ even though the folding
times got longer. On the other hand, below $T_{min}$, this process becomes
more extended in time.

\begin{figure}
\epsfxsize=3.4in
\centerline{\epsffile{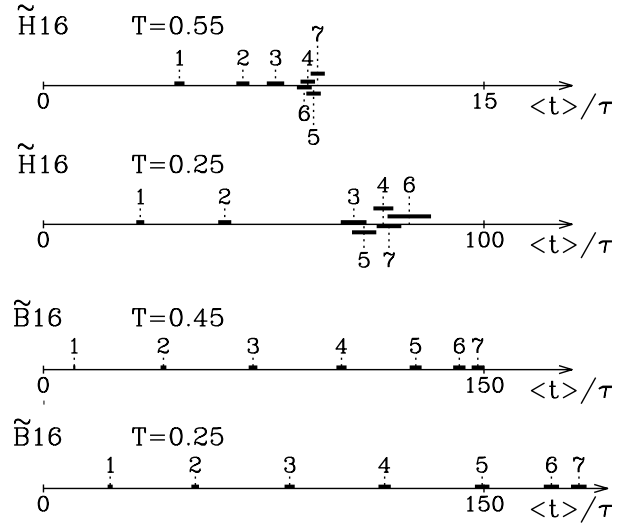}}
\caption{The contact sequencing for sequences \~H16 and \~B16 at
temperatures that are above or below $T_{min}$.
}
\end{figure}

\subsection{Examples of the trajectories}

Notice that the first appearance of the last contact that is
monitored does not coincide yet with the full folding because
a substantial time is still needed to lock precisely into the native basin.
In fact, getting to the stage where the last contact starts giving
contribution to the energy takes of order of only from 5\% to 10\% of the full
folding time.  This time roughly corresponds to the collapse of
the chain. 

This is illustrated in Figures 13 and 14, for \~H16 and \~B16 respectively,
where examples of single trajectories at $T_{min}$ are shown.
The trajectories are characterized by their total energy, $E$, potential 
energy, $E_p$, the radius of gyration, $R_g$, conformational distance away from
the native state and the number of all native contacts, $N_c$, 
(not only of those few that were discussed in the context of sequencing)
present.
All of these parameters depend on time essentially monotonically.
The establishment of the short contacts takes place rapidly and then
the system spends most of the folding time ``floating'' near the
native basin until it really finds it. Thus what takes most of the time in 
folding is the process of search. This process is thus governed more by
kinetics than by energetics.

\begin{figure}
\epsfxsize=3.2in
\centerline{\epsffile{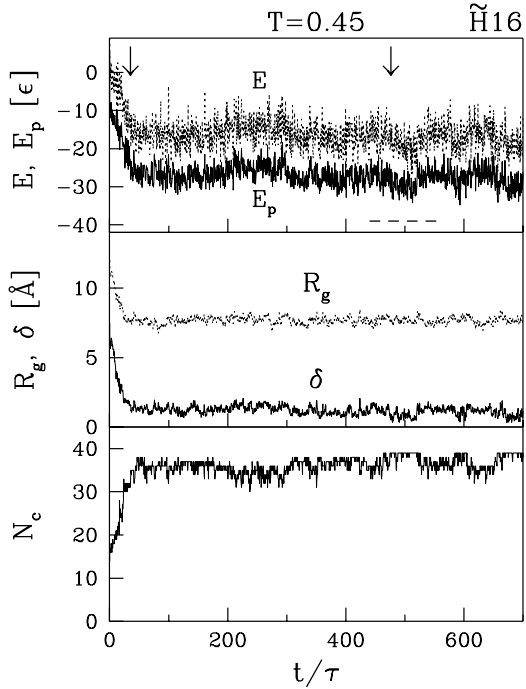}}
\caption{A typical folding trajectory at $T_{min}$ for sequence \~H16.
In the top panel the left arrow indicates when the last contact appears
for the first time, and the right arrow indicates when folding
takes place, i.e. $\delta < \delta_c$. The horizontal dash line indicates
the energy of the native state.}
\end{figure}

\begin{figure}
\epsfxsize=3.2in
\centerline{\epsffile{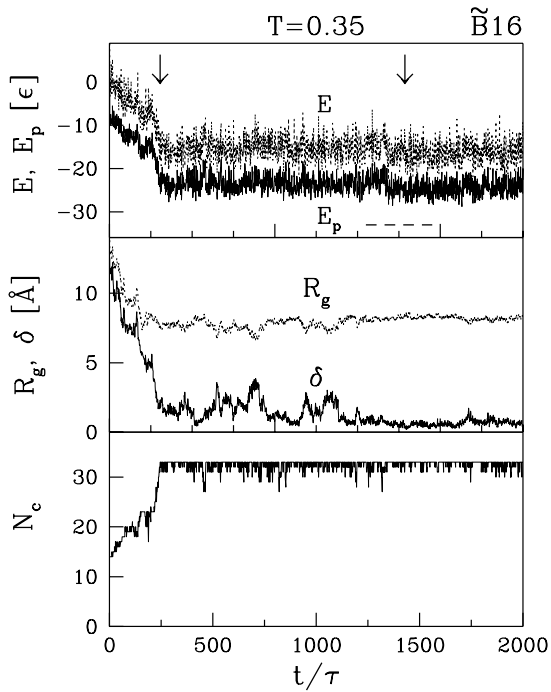}}
\caption{Same as Figure 12 but for sequence \~B16.}
\end{figure}

\subsection{Dependence on the viscous friction}

We now examine how do the folding properties depend on the viscosity.  So
far, the results of all of the simulations shown here were obtained with
$\gamma=10 m/\tau$.  We have checked that for sequence H16, $T_{min}$ does
not depend on the friction and is equal to 0.3 if $\gamma > 0.2 m/\tau$. At
very low  friction, i.e. when $\gamma < 0.2 m/\tau$, we observe some
decrease of $T_{min}$. For instance, $T_{min}$ is about 0.05 for
$\gamma=0.005 m/\tau$.  Since $T_f$ cannot depend on $\gamma$, at this very
low value of friction the sequence becomes an excellent folder. However,
such small values of $\gamma$ are very unrealistic for
proteins.\cite{Thirumalai,Klimov1} We have also checked that other
thermodynamic properties do not depend on $\gamma$.  Figure 15 shows the
dependence of the median folding time on friction for a wide range of
$\gamma$, at $T=0.3$ and also at $T_{min}$ for $\gamma < 0.2 m/\tau$.
Notice that there is a minimum of the folding time at some intermediate
friction.  The folding time grows not only at higher friction, as $\gamma$
increases, but also at very low friction, as $\gamma$ decreases towards 0.
There is no folding at $\gamma=0$, when the total energy of the system is
conserved  -- a coupling to a heat reservoir is essential for folding.  The
optimal value of $\gamma$, at which folding is the fastest, is in a range
of from 0.2 to 0.5. This observation is in agreement with the results of
Klimov and Thirumalai on the viscosity dependence of the folding
rate.\cite{Klimov1}  Notice that Klimov and Thirumalai have determined
foldability rates and not the folding times themselves.  We also observe
that for $\gamma$ between 1 and 30, the folding time of sequence H16 grows
almost linearly with $\gamma$, as shown in the inset of Figure 15.  By
extrapolation we estimate that for $\gamma=50 m/\tau$ the folding time for
H16 would be $1500\tau$ (for \~H16 it would be about $3000\tau$).
Veitshans {\it et al.}\cite{Thirumalai} have estimated that at this
apparently more realistic value of friction, corresponding to the case of
an single aminoacid in a water solution at room temperature, the time unit
of $\tau$ should be about $3ns$.  Hence our estimate of the folding time of
a 16-monomer $\alpha$-helix is from 4 to 10 $\mu s$. If we assume that for
the $\beta$-structures the folding time also scales linearly with $\gamma$,
then the time scale for the formation of a 16-monomer $\beta$-hairpin would
range from 24 to 30 $\mu s$. For comparison, experimental results have
shown that a $\beta$-hairpin consisting of 16 aminoacids can fold as fast
as $6\mu s$ at room temperature.\cite{Munoz} An $\alpha$-helix of the same
length folds 30 times faster. Notice that our estimates yield larger
folding times than observed in experiments but the orders of magnitude
agree. 

\begin{figure}
\epsfxsize=3.2in
\centerline{\epsffile{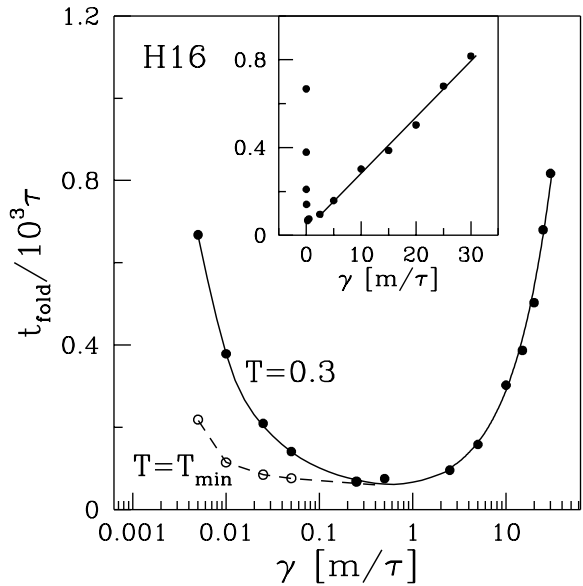}}
\caption{The median folding time at $T=0.3$ (continuous line)
as function of the friction coefficient, $\gamma$, for sequence H16.
For $\gamma < 0.2 m/\tau$, $T_{min}$ becomes smaller than 0.3 and the folding
times at $T=T_{min}$ are shown for several values of $\gamma$
(discontinuous line). For $\gamma > 0.2 m/\tau$, $T_{min}$ is equal to 0.3.
The points are based on 200 trajectories.
The inset shows the dependence at $T=0.3$ in a non-logarithmic scale of
$\gamma$, where for the higher values of $\gamma$, the points are fitted by a
straight line. }
\end{figure}

\subsection{Folding with one end fixed}

Finally, we examine a special case, when one end of the chain is fixed
during folding. This is relevant to the process of protein synthesis: one
end can be thought of as being momentarily glued to the surface of a
ribosome until the whole protein is constructed by adding new
segments.\cite{Streyer} In the synthesis process, the protein folds as it
is being produced and its length extends. At each instant, however, one end
of the protein can be considered pinned.

Figure 16 shows that the helix H16 with the end bead  fixed has a
considerably lower value of $T_{min}$, compared to no clamping, and it
generally folds faster at low temperatures.  pinned folds faster below
$T_{min}$ and is characterized by a significantly lowered $T_{min}$.  We
have also found that $T_f$ is not affected by the pinning.  Thus the helix
becomes a much better folder.  On the other hand, as seen in Figure 16, the
contact sequencing becomes somewhat disturbed.  Folding starts at the
unclamped end.  After the initial stage, many contacts get established
almost simultaneously in an avalanche-like process.  Then there is a large
gap in time before the last contact, near the clamped end, comes into
place.

Analyzing the phononic spectrum of the pinned helix H16 yields 
$\;\omega_1 \tau \approx 1.964$ for the first excited mode
which is almost the same as for H16 without the pinning.

For the $\beta$-sequences the different scenario is different: the pinned
$\beta$-structures seem to fold slower than the unpinned ones. For
instance, fixing one end of sequence B16 increase the folding time by about
60\%, at $T=0.05$. The pinned B10 also folds slower, but not significantly
slower.

Our results indicate that the process of protein synthesis itself may
accelerate the folding process when a helix part is being produced.  Short
$\beta$-structures may introduce some slowing down.

\begin{figure}
\epsfxsize=3.4in
\centerline{\epsffile{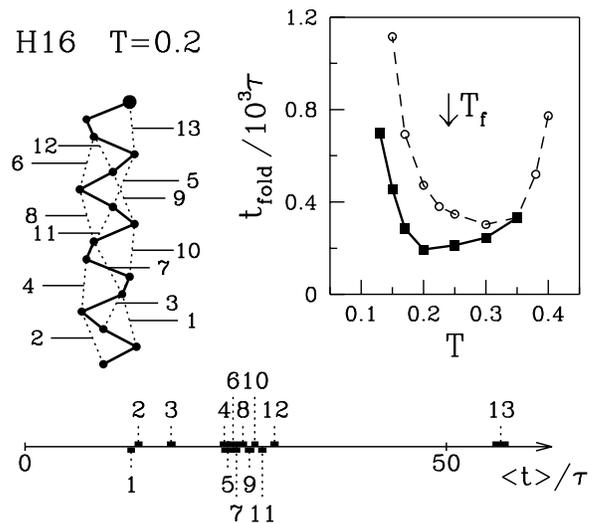}}
\caption{The order of contact appearance together with the average times
for the contacts to be established for the first time for sequence H16 with
one end monomer fixed. The fixed monomer is shown enlarged. The results are
averaged over 1000 trajectories at $T_{min}$.  The inset compares the
temperature dependencies of the median folding time for the situation in
which the end monomer is (solid line) or is not (broken line) clamped.}
\end{figure}

\section{CONCLUSIONS}

In summary, we have studied the Go-type models of off-lattice secondary
structures of proteins.  We have provided a systematic characterization of
both equilibrium and kinetic properties of these models.  Models with the
steric constraints were endowed with better thermodynamic stability. The
stability can be assessed from the phononic spectra.  The folding times
strongly depend on the system size and on the geometry of the native state.
The $\alpha$-helices, which are stabilized merely by the local contacts
appear to have much better folding properties than the $\beta$-structures.
However, it should be noted that for both kinds of the structures none of
the models is a very good folder -- $T_f$ is always smaller than $T_{min}$,
-- although the choice of the native basin size $\delta_c$ has been made
consistently by the shape distortion technique. We have checked that if one
uses a somewhat larger $\delta _c$ then this results in an increase in the
effective values of both $T_f$ and $T_{min}$  and thus their ordering
remains unchanged.  The kinetic criterion for good foldability does not
appear to be compatible with the thermodynamic criterion provided by
Klimov and Thirumalai \cite{Klimov2} in the case of the
$\beta$-sequences.  We speculate that this may be related to the low
compactness level of their native structures.
Folding of the secondary structures at $T_{min}$ proceeds, on average, 
through a well defined sequence of events.
That sequencing may become more complicated in proteins when
several secondary structures compete when folding.  
We intend to explore this issue later.

\section{ACKNOWLEDGMENTS}

The idea of this project arose in discussions with Jayanth R. Banavar,
whose subsequent interest and encouragement were vital for its completion.
Many discussions with M. S. Li are also appreciated. We also thank M. Geller
for pointing out to us that the folding process may 
be affected by the pinning.
This work was supported by KBN (Grant No. 2P03B-025-13).

\end{document}